\documentclass[twocolumn,showpacs,superscriptaddress,reprint,floatfix,aps,prl]{revtex4-1}

\usepackage{amsmath,amssymb,amsthm}
\usepackage{graphicx}
\usepackage{lineno}

\usepackage{hyperref}
\hypersetup{
    colorlinks=true,
    linkcolor=cyan,
    filecolor=magenta,      
    urlcolor=blue,
    citecolor=blue,
}

\usepackage{cleveref}
\usepackage{comment}

\usepackage{graphicx}
\usepackage{amsmath}

\usepackage{dcolumn}
\usepackage{bm}

\usepackage{array}

\begin{document}

\title{Swimming, Feeding and Inversion of 
Multicellular Choanoflagellate Sheets}

\author{Lloyd Fung}
\email[]{lsf27@cam.ac.uk}
\affiliation{Department of Applied Mathematics and Theoretical 
Physics, Centre for Mathematical Sciences,\\ University of Cambridge, Wilberforce Road, Cambridge CB3 0WA, 
United Kingdom}
\author{Adam Konkol}
\email[]{ak2351@cam.ac.uk}
\affiliation{Department of Applied Mathematics and Theoretical 
Physics, Centre for Mathematical Sciences,\\ University of Cambridge, Wilberforce Road, Cambridge CB3 0WA, 
United Kingdom}
\author{Takuji Ishikawa}
\email[]{t.ishikawa@tohoku.ac.jp}
\affiliation{Department of Biomedical Engineering, Tohoku University, 6-6-01 Aoba, Aramaki, Aoba-ku, Sendai 980-8579, Japan}
\author{Ben T. Larson}
\email[]{blarson@berkeley.edu}
\affiliation{Department of Biochemistry \& Biophysics, 
University of California, San Francisco, 600 16th St., San Francisco, CA  94143-2200, USA}
\author{\\Thibaut Brunet}
\email[]{thibaut.brunet@pasteur.fr}
\affiliation{Department of Cell Biology and Infection, Institut Pasteur, 25-28 rue du Dr. Roux, 75724 Paris Cedex 15, France}
\author{Raymond E. Goldstein}
\email[]{R.E.Goldstein@damtp.cam.ac.uk}
\affiliation{Department of Applied Mathematics and Theoretical 
Physics, Centre for Mathematical Sciences,\\ University of Cambridge, Wilberforce Road, Cambridge CB3 0WA, 
United Kingdom}%

\date{\today}

\begin{abstract}
The recent discovery of the striking sheet-like multicellular choanoflagellate species 
{\it Choanoeca flexa} that dynamically interconverts 
between two hemispherical forms of opposite orientation raises fundamental questions in
cell and evolutionary biology, as choanoflagellates are the closest living relatives of animals.  
It similarly motivates questions in fluid and solid mechanics concerning the differential swimming
speeds in the two states and the mechanism of curvature inversion triggered by
changes in the geometry of microvilli emanating from each cell.  Here we develop fluid dynamical 
and mechanical models to address these observations and show that they capture the main features of 
the swimming, feeding, and inversion of {\it C. flexa} colonies.

\end{abstract}
\maketitle

Some of the most fascinating processes in the
developmental biology of complex multicellular organisms involve radical changes in
geometry or topology.  From the folding of
tissues during gastrulation \cite{gastrulation} to 
the formation of hollow spaces in plants \cite{Goriely}, these
processes generally involve coordinated cell shape
changes, cellular division, migration and apoptosis, and 
formation of an extracellular matrix (ECM).
It has become clear through multiple strands of
research that evolutionary precedents for
these processes exist in some of the 
simplest multicellular organisms such as green 
algae \cite{hohn2016distinct,Volvox_prl} and choanoflagellates \cite{BrunetKing}, 
the latter being the closest living
relatives of animals.  Named for their 
funnel-shaped collar of microvilli that facilitates
filter feeding from the flows driven by their 
beating flagellum, choanoflagellates serve as model organisms
for the study of the evolution of multicellularity.

\begin{figure}[b]
\centering
	\includegraphics[trim={0 0cm 0 0cm}, clip, width=0.98\columnwidth]{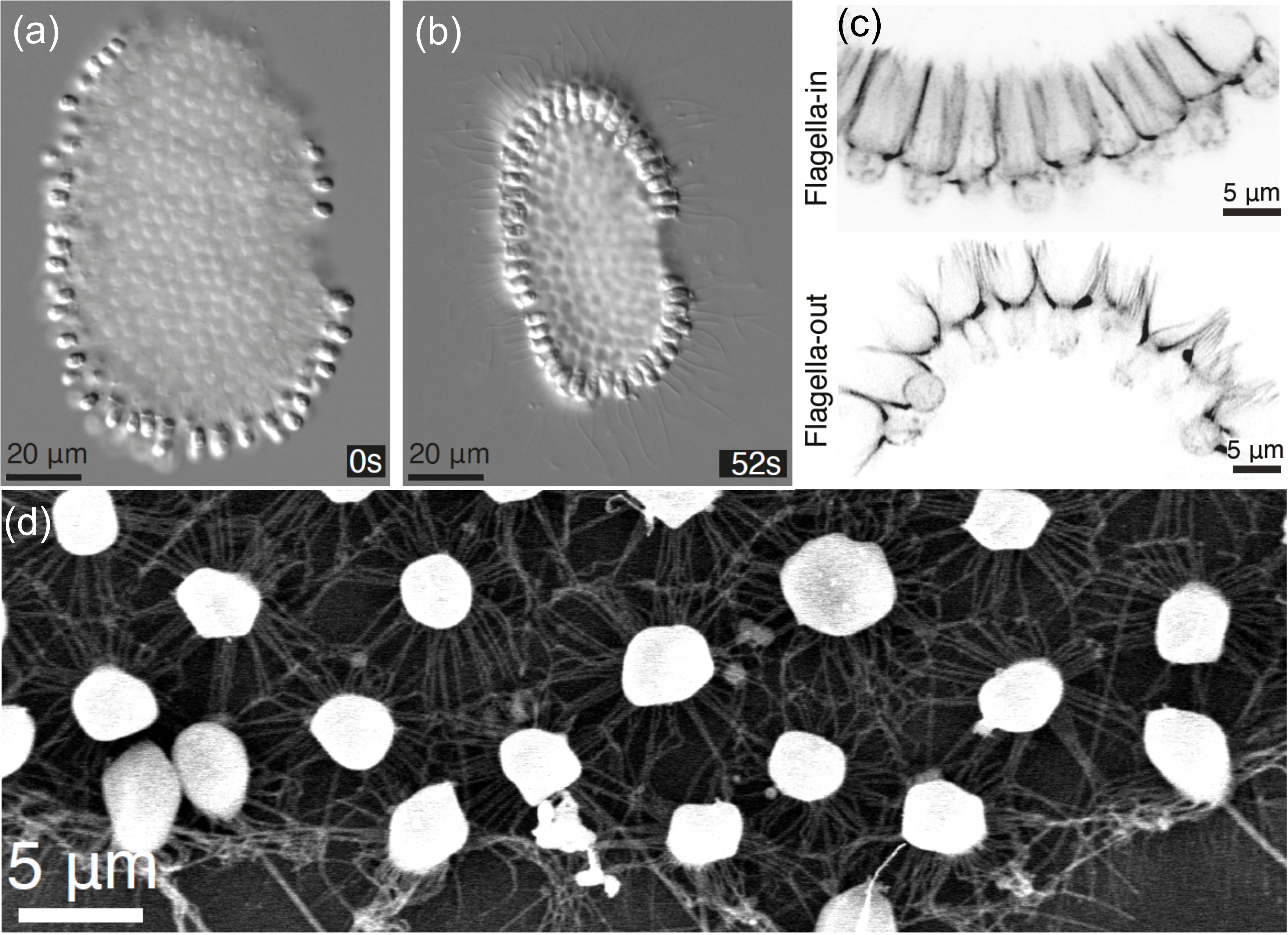}
	\caption{The multicellular choanoflagellate {\it Choanoeca flexa}. Top views of (a) {\it flag-in} and (b) {\it flage-out}
	states at times relative to removal of light. (c) Close-up of the collar connections in the two states. 
	(d) Electron micrograph showing round white cell bodies connected by microvilli.  Adapted from 
	\cite{brunet2019light}.}
	\label{fig1}
\end{figure}

While well-known multicellular choanoflagellates exist as linear chains or ``rosettes"
\cite{fairclough2010multicellular} held together by an ECM \cite{larson2020biophysical}, a 
new species named {\it Choanoeca flexa} was recently discovered 
\cite{brunet2019light} with an unusual
sheet-like geometry (Fig. \ref{fig1}) in which hundreds of cells adhere to each 
other by the tips of their microvilli, without an ECM \cite{leadbeater1983life}. 
The sheets can exist in two forms with opposite curvature, one with
flagella pointing towards the center of curvature [``{\it flag-in}"] with a relatively
large spacing between cells, and another with 
the opposite arrangement [``{\it flag-out}"] with more tightly-packed cells.
Transformations from {\it flag-in} to {\it flag-out} can be triggered by darkness,
and occur in $\sim \! 10\,$s.  
Compared to the {\it flag-out} form, 
the {\it flag-in} state has limited motility and is better suited to 
filter-feeding.  It was conjectured \cite{brunet2019light} that
the darkness-induced transition to the more motile form is a type of photokinesis. 

 \begin{figure*}[t]
	\includegraphics[width= .93\textwidth]{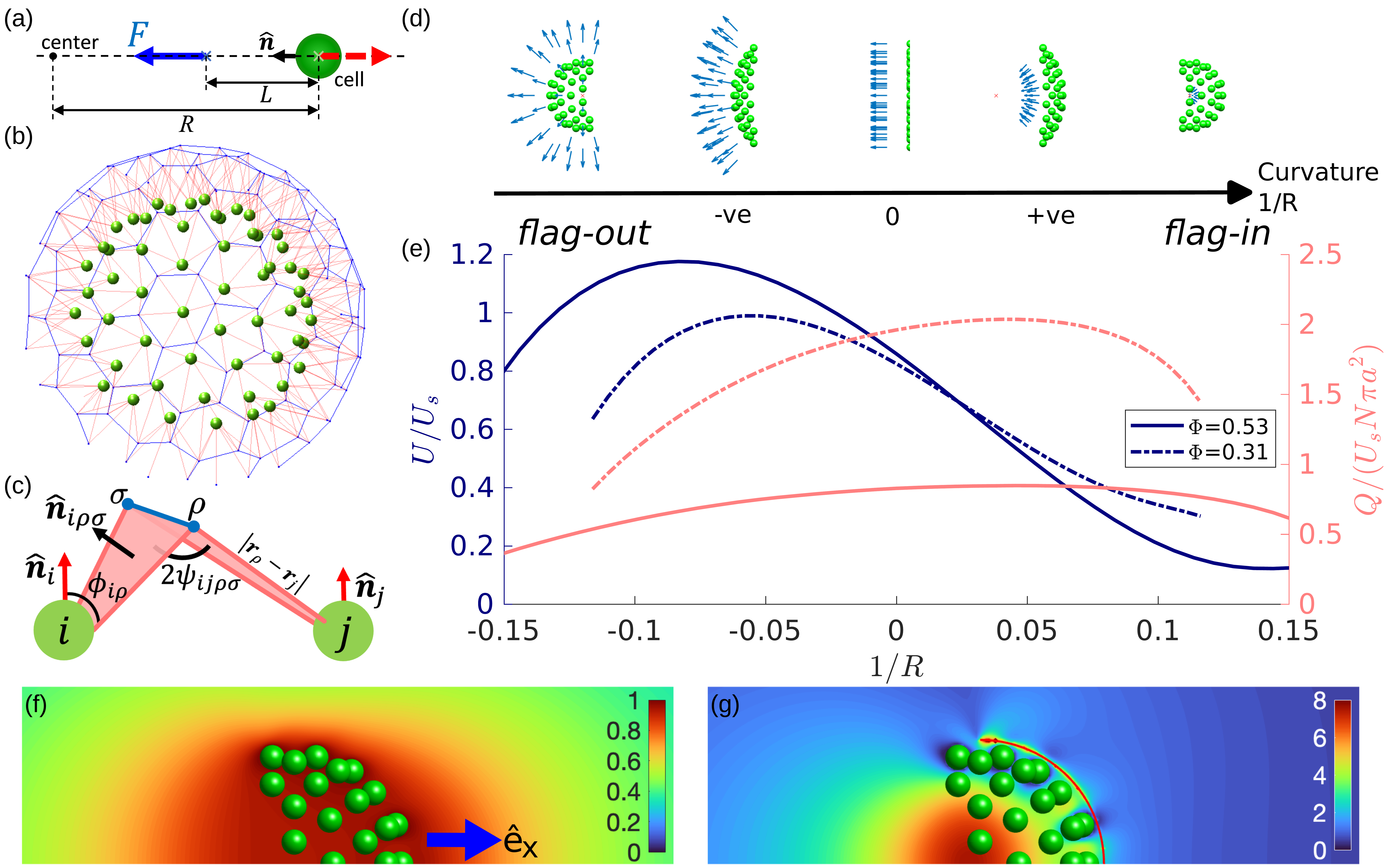}
	\caption{Models for \emph{C. flexa}. 
	(a) Cell body and flagellar force in the \textit{flag-in} state. 
	(b,c) Mechanical model of interconnecting microvilli in rafts; cells (green spheres, not to scale) are at the vertices of a geodesic icosahedron.  Blue arrows indicate flagella forces, 
	red segments represent microvilli, 
	blue dots the microvilli tips, and blue lines the collar-collar interface. (b) Connectivity of the whole raft. (c) Connections between two cells.
    Effect of curvature $1/R$ on (d) geometry of raft and (e) swimming speed $U$ (blue) and flow rate $Q$ (red) passing through $S_f$ at constant.
    (f,g) Cross section of the disturbance flows $\tilde{\mathbf{u}}_{d}$ and $\mathbf{u}_f$ around the a raft ($\Phi=0.31$, $R=8.61$) in the reciprocal problems for calculating $U$ and $Q$.
	}
	\label{fig2}
\end{figure*}

As a first step toward understanding principles that govern the behavior of such 
a novel organism as {\it C. flexa}, we analyze two models for these shape-shifting structures.  
First, the fluid mechanics are studied by 
representating the cell raft as a collection of spheres   
distributed on a hemispherical surface, with nearby point forces 
to represent the action of flagella.
Such a model has been used to describe the motility of small sheet-like 
multicellular assembles such as the alga {\it Gonium} \cite{GoniumPRE}.  The 
motility and filtering flow through these rafts as a function of cell spacing and curvature 
explain the observed properties of {\it C. flexa}.  Second, abstracting the complex
elastic interactions between cells to the simplest connectivity, we show that a
model based on linear elasticity at the microscale produces bistability on the colony scale.

\emph{Fluid mechanics of feeding and swimming} --  The cells in a {\it C. flexa} raft are 
ellipsoidal, with major and minor axes $a\sim 4\,\mu$m and 
$b\sim 3\,\mu$m, with a single flagellum of length $2L\sim 24.5\,\mu$m 
and radius $r\sim 0.5\,\mu$m beating with amplitude $d\sim 2.3\,\mu$m and 
frequency $f\sim\! 44\,$Hz \cite{SM}, sending bending waves away from the body.
A cell swims with flagellum and collar rearward; 
the body and flagellum comprise a ``pusher" force dipole.  From resistive force theory \cite{Laugabook} we estimate the 
flagellar propulsive
force to be $F\sim 2L\left(\zeta_\perp-\zeta_\parallel\right) (1-\beta)  f \lambda \sim 8\,$pN, where $\beta$ is a function of the wave geometry, $\lambda \sim 12\, \mu$m is the wavelengths along the direction of the bending wave \cite{SM}, $\zeta_\perp$ and 
$\zeta_\parallel$ are transverse and longitudinal drag coefficients, 
$\zeta_\perp\sim 2\zeta_\parallel\sim 4\pi\mu/\ln(2L/r)$, with $\mu$ the fluid viscosity.  
These features motivate a computational model in which 
$N$ identical cells in a raft have a spherical 
body of radius $a$ and a point force $F\hat{\bf n}_i$ acting on the fluid a distance $L$ from the sphere center, oriented along 
the vector $\hat{\bf n}_i$ that represents the collar axis
[Fig. \ref{fig2}(a,b)].
An idealization of a curved raft involves placing those spheres on a connected subset of the 
vertices of a {\it geodesic} icosahedron (one whose vertices lie on a spherical surface) of radius $\rho\gg a$;
the area fraction $\Phi$ 
of the sheet occupied by cells scales as $\Phi\sim N(a/\rho)^2$.
The pentagonal neighborhoods within the geodesic icosahedron serve as topological defects that allow for smooth large-scale
surface curvature \cite{seung1988defects}.  Importantly, 
confocal imaging of
{\it C. flexa} colonies shows that a significant fraction ($\sim\!\!0.25$) 
of the cellular neighborhoods defined by the microvilli connections 
are pentagonal \cite{SM}, and earlier work on {\it C. perplexa} \cite{leadbeater1983life} also found non-hexagonal packing.
We use the geodesic icosahedron $\{3,5+\}_{(3,0)}$ in standard notation \cite{Wenninger},   
with $92$ total vertices, and take patches with $N=58$ for computational tractability. The vectors $\hat{\bf n}_i$ point towards (away from) the 
icosahedron center in the 
{\it flag-in} ({\it flag-out}) forms (Fig. \ref{fig2}(a)). 
A deformation of the sheet to a new radius 
$\rho'$, at fixed $\Phi$, requires the new polar angle $\theta_i^\prime$ of a 
cell with respect to the central axis of the sheet be related to its original angle $\theta_i$ via
${\rho^\prime}^2\left(1-\cos{\theta_i}^\prime\right)= \rho^2\left(1-\cos\theta_i\right)$.
We define the scaled 
force offset length $\ell=L/ a\sim 3$ and sheet radius $R=\rho^{\prime}/a \gtrsim 6$, and take $R>0$ in the {\it flag-in}
state.

Images of many colonies of {\it C. flexa} \cite{SM} show that 
the packing
fraction in the
{\it flag-out} state $\Phi_{\rm out}=0.47\pm 0.06$, considerably
less than both the maximum packing fraction $\Phi_{\rm max}=\pi\sqrt{3}/6\simeq 0.907$ 
for a hexagonal array of spheres in a plane, and the estimated maximum packing fraction
$\tilde{\Phi}_{\rm max}\simeq 0.83$ for circles on a sphere \cite{circlepacking}.
The packing fraction in the {\it flag-in} state is $\Phi_{\rm in}=0.34\pm 0.03$, and we use the 
extremes $\Phi_{\rm out}=0.53$ and $\Phi_{\rm in}=0.31$ as representative 
values to explore the consequences of the differences between the two forms. 

Consider first an isolated force-free spherical
cell at the origin moving at velocity $U_s\hat{\mathbf{e}}_x$ with point force $-{\bf F}=-F \hat{\mathbf{e}}_x$ at $-L\hat{\mathbf{e}}_x$ acting on the fluid
and its reaction force ${\bf F}$ acting on the cell. 
The cell experiences Stokes drag $-\zeta_s U_s \hat{\mathbf{e}}_x$, where $\zeta_s=6\pi\mu a$, 
and a disturbance drag $-D \hat{\mathbf{e}}_x$ arising from the disturbance flow created by the point force.  By the reciprocal theorem \cite{RT}, the disturbance drag is ${D}={\bf F} \cdot \tilde{\mathbf{u}}_d(-L \hat{\mathbf{e}}_x)/\hat{U}$, where $\tilde{\mathbf{u}}_{d}({\bf r})$ is
 the disturbance flow created when the cell is dragged along $\hat{\mathbf{e}}_x$ with unit speed $\hat{U}$.  Force balance then yields the single-cell swimming speed $U_s \equiv (F/\zeta_s)[1-3/(2\ell)+1/(2\ell^3)]$. 
Thus, the closer the point force is to the cell 
(i.e., the smaller is $\ell$), the more drag the cell experiences and the slower is ${U}_s$.  
Setting $\ell = 3$  yields $U_s\sim 118\, \mu$m/s, consistent with observations.

This intuitive picture extends to a raft of cells. As the raft moves at velocity $U \hat{\mathbf{e}}_x$, it experiences a Stokes drag 
$-\zeta U\hat{\mathbf{e}}_x$. The disturbance flow created by the point forces $F \hat{\bf n}_i$ acting at $\mathbf{r}_i+L \hat{\mathbf{n}_i}$, produces a disturbance drag $D=F\Sigma_i \hat{\bf n}_i \cdot \mathbf{u}_d(\mathbf{r}_i+L\hat{\mathbf{n}_i})$, where ${\mathbf{u}}_{d}$ is the 
(dimensionless) disturbance flow from the raft when it is dragged along $\hat{\mathbf{e}}_x$ with unit speed. 
Force balance then yields 
\begin{equation}
  U=-\frac{F}{\zeta} \sum_{i=1}^N \hat{\bf n}_i \cdot \left[ \hat{\mathbf{e}}_x- 
  {\mathbf{u}}_{d}\left(\mathbf{r}_i + L \hat{\bf n}_i\right) \right], \label{eq:motility_x}
\end{equation}
where  $-F\Sigma_i\hat{\bf n}_i$ is the sum of reaction forces propelling the raft along $\hat{\mathbf{e}}_x$,
and where ${\mathbf{u}}_{d}$ has been rendered dimensionless by the unit speed.
In practice, we compute ${\mathbf{u}}_{d}$ and $\zeta$ using a Boundary Element Method \cite{GoniumPRE}.
Because of the curved geometry, point forces are closer to neighboring cells in the {\it flag-in} state than in the {\it flag-out} state. 
Thus, as in Fig. \ref{fig2}(d), for a geometry with a given $|R|$, the 
{\it flag-in} state has a larger disturbance drag than the {\it flag-out} state, and a smaller speed $U$.

The difference in swimming speed between the two states can also be explained in terms of ${\mathbf{u}}_{d}(\mathbf{r}_i + L \hat{\bf n}_i)$ in (\ref{eq:motility_x}).
Figure \ref{fig2}(f) shows that ${\mathbf{u}}_{d}$ inside the raft is close to $\hat{\mathbf{e}}_x$ because of the curved geometry and screening effects.
Hence, $U$ is small when the point forces are inside. 
Meanwhile, ${\mathbf{u}}_{d}$ outside decays with the distance from the raft, so $U$ is large when $\mathbf{r}_i + L \hat{\bf n}_i$ is outside.

Previous work on filter-feeding in choanoflagellates focused first on the far-field limit based on a 
stresslet description \cite{Roper}, but later work showed near-field
effects can significantly affect capture rates \cite{Kirkegaard}.
To estimate the filter-feeding flux $Q$ passing through a colony of {\it C. flexa}, we measure, in the body frame, the flux passing through the surface $S_f$ projected a distance of $1.2 a$ from the cell center along $\hat{\bf n}$, as in Fig. \ref{fig2}(g).
By the reciprocal theorem, $Q$ can be written in terms of the disturbance flow ${\mathbf{u}}_{f}$ around a stationary raft and the hydrodynamic forces ${\mathbf{F}}_f$ on the raft when the surface $S_f$ applies a unit normal pressure $\hat{\mathbf{p}}$ on the fluid,
\begin{equation}
  Q = F\sum_i \hat{\mathbf{n}}_i \cdot {\mathbf{u}}_{f} (\mathbf{r}_i + L \hat{\bf n}_i) + {U} \hat{\mathbf{e}}_x \cdot\left({\mathbf{F}}_f -\int_{S_f} \hat{\mathbf{p}} \, dA\right), \label{eq:body_flux}
\end{equation}
where ${\mathbf{u}}_{f}$ and ${\mathbf{F}}_f$ acquire the units of velocity/pressure and area, respectively,
by scaling with $\vert \hat{\mathbf{p}}\vert$.
Numerical results in Fig. \ref{fig2}(d) show that the flux due to point forces $\sum_i F \hat{\mathbf{n}}_i \cdot {\mathbf{u}}_{f}/\hat{p}$ strongly dominates $Q$.
Therefore, the difference in $Q$ between the two states can be explained by ${\mathbf{u}}_{f}(\mathbf{r}_i + L \hat{\bf n}_i)$ (Fig. \ref{fig2}(g)).
To maintain incompressibility under pressure $\hat{\mathbf{p}}$, the disturbance flow ${\mathbf{u}}_{f}$ is much stronger inside the raft than outside.
Hence, point forces placed inside the raft pump more flow through the raft than when placed outside. 

Figure \ref{fig2}(d) shows the effect 
of changes in the raft curvature and packing 
fraction.  There is one $R$ that maximizes 
swimming speed in the {\it flag-in} state and another one that maximizes feeding
flux in the {\it flag-out} state. This arises from a balance between 
the screening effect mentioned above and the alignment of forcing. In the 
{\it flag-out} state, an initial decrease in curvature aligns the forcing direction 
with the swimming direction, increasing swimming speed, but a 
further reduction in curvature reduces the screening effect as cells are 
now more spread out in the plane orthogonal to the swimming direction. A 
similar argument applies to the flow rate maximum in the {\it flag-in} 
state. Comparing these maxima, 
Fig. \ref{fig2}(d) shows that a spread-out colony results in more flux, while a closely-packed colony results in 
faster motility.  
Thus, through the interconversion between the two states, 
{\it C. flexa} takes advantage of the hydrodynamics effect of the curved geometry for efficient filter-feeding and swimming. 

\begin{figure*}[t]
    \centering
   \includegraphics[width=0.83\textwidth]{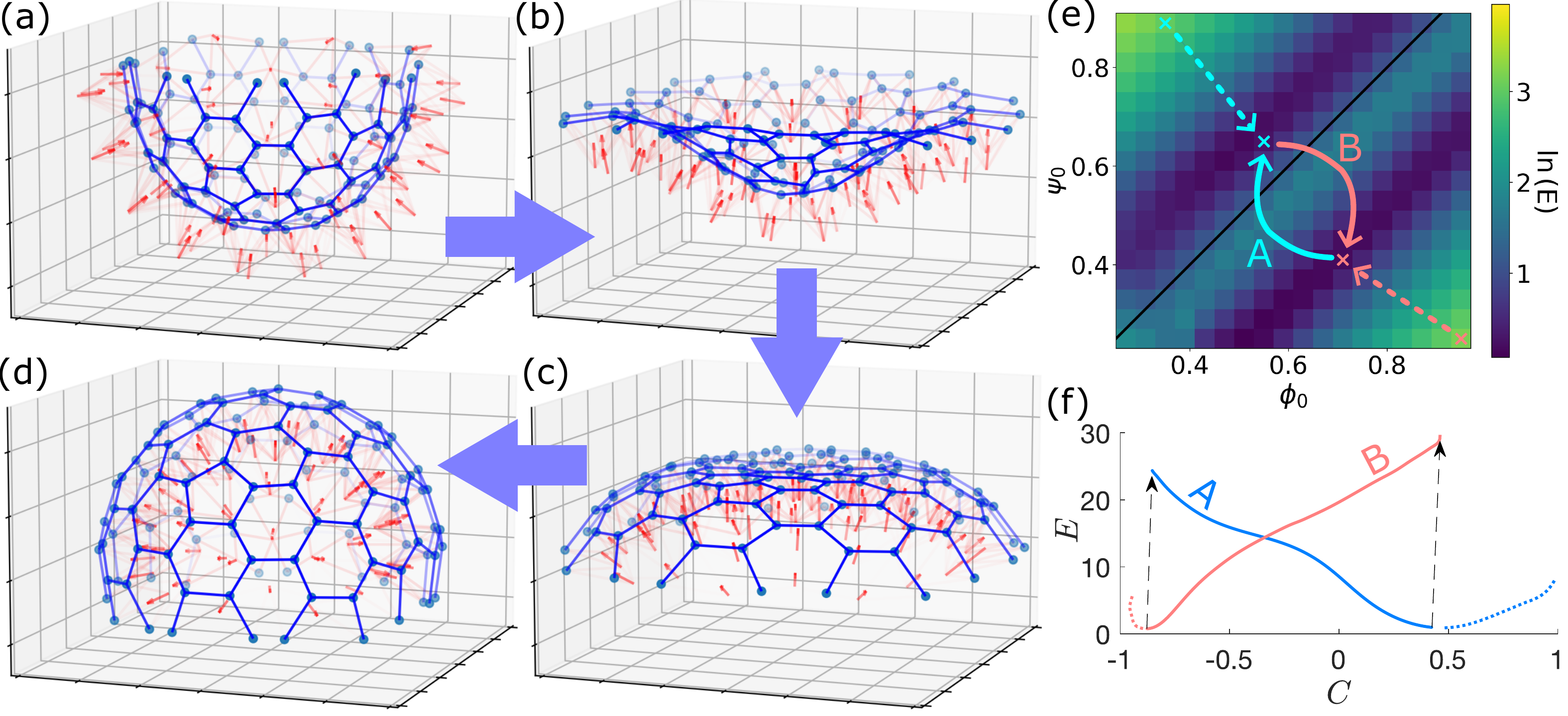} 
    \caption{Inversion dynamics from numerical studies.  (a)-(d) A colony, initially at
    a hemispherical minimum with $(\phi_0,\psi_0)=(0.55,0.65)$, inverts after a change 
    to $(0.71,0.41)$, with $\ell_0=0.5$, $K_\psi=2$ and $K_\ell=5$.   Connections between collar vertices are shown in blue, apicobasal axes as red arrows at cell body positions. (e) Minimum energy $E$ in the $\psi_0-\phi_0$ plane, where $C<0$ ({\it flag-in}) 
    above the black line $\psi_0=\phi_0$, and $C > 0$ ({\it flag-out}) below. $(f)$ Evolution of $E$ vs $C$ as the colony relax towards a minimum energy state after instantaneous changes in $(\phi_0,\psi_0)$ shown by the dotted and solid red and blue lines in $(e)$.}
    \label{fig4}
\end{figure*}

\emph{Mechanics of inversion} -- 
Detailed studies suggest that inversion requires an active 
process within each cell, likely driven by contraction of an F-actin ring at the
apical pole through the action of myosin \cite{brunet2019light}.
Thus, a full treatment would address the complex problem of 
elastic 
filaments responding to the apical 
actomyosin system and adhering to each other. 
We simplify this description by considering as in Fig. \ref{fig2}(e) that each cell $i$, located at ${\bf r}_i$ and surrounded by $m_i$ 
neighbors, has $m_i$ rigid, straight filaments 
emanating from it.  Two filaments from neighboring cells $i$ and $j$ meet at vertex $\rho$ located at ${\bf r}_\rho$, with $\phi_{i\rho}$
the angle between ${\bf r}_\rho-{\bf r}_i$ and the cell normal 
vector $\hat{\bf n}_i$.  
Any two adjacent
filaments emanating from cell $i$, and which meet neighboring
filaments at vertices $\rho$ and $\sigma$, define a plane whose normal 
$\hat{\bf n}_{i \rho \sigma}$ points toward the apicobasal axis $\hat{\mathbf{n}}_i$. 
That
normal and its counterpart $\hat{\bf n}_{j \sigma \rho }$
on cell $j$ determine the angle $2\psi_{ij\rho\sigma}$
between the two planes.

As above, we use the geodesic icosahedron to define the cell positions and 
thus determine the filament network connecting neighboring cells.  The
two sets of angles $\{\phi\}$
and $\{\psi\}$ are used to define a Hookean elastic energy that 
mimics the elasticity of the microvilli, allowing for 
preferred intrinsic angles $\phi_0$ and $\psi_0$ that encode the effects of the apical actomyosin system on the microvilli and the geometry of
microvilli adhesion.  Allowing also for stretching away from a rest length $\ell_0$, the energy is
\begin{align}
    E = \frac{1}{2}k_\phi\sum_{i,\rho} \delta\phi_{i\rho}^2 
    + \frac{1}{2}k_\psi\!\!\sum_{i,j,\rho,\sigma}\!\! \delta\psi_{ij\rho\sigma}^2
    + \frac{1}{2}k_\ell\sum_{i,\rho} \delta\ell_{i\rho}^2, \label{eq:mech_e}
\end{align}
where $\delta\phi_{i\rho}=\phi_{i\rho}-\phi_0$, $\delta\psi_{ij\rho\sigma}=\psi_{ij\rho\sigma}-\psi_0$, and $\ell_{i\rho}=\vert {\bf r}_i-{\bf r}_\rho\vert-\ell_0$.  The 
moduli $k_\phi$, $k_\psi$ and $k_\ell$ and quantities
$\phi_0,\psi_0$ and $\ell_0$ are assumed constant for all cells. 

The energy \eqref{eq:mech_e} is intimately tied to the lattice geometry of the raft.
If the cells are arranged in a hexagonal lattice ($m_i=6$) the system of 
filaments can achieve $E=0$ by setting all cell-collar angles to $\phi_0$, all collar-collar interface angles to $\psi_0$, and $\phi_0=\psi_0$.
This corresponds to a flat sheet.  Increasing $\phi_0=\psi_0$ leads to uniform, isotropic sheet expansion.
In a non-planar raft, curvature is introduced through topological defects 
($m_i \neq 6$), such as pentagons, and mismatch between the local values of $\psi$ and $\phi$.  While pentagonal defects are known to 
cause out-of-plane buckling in crystal lattices \cite{seung1988defects},  
they do not by themselves select a particular {\it sign} of the induced curvature.
Thus, there is inherent
bistability in the cellular raft that can be biased by changes in the
geometry of the out-of-plane filaments, somewhat akin to the role of
``apical constriction" in the shapes of epithelia \cite{Hannezo}.

For the case of two cells lying in a plane, each with two filaments, 
and with one vertex between them,
if $\phi=\phi_0$, $\psi=\psi_0$, and $r=\ell_0$, then the filament tips
lie on a circle of radius 
$R_0=1/C_0$, where $C_0=\sin(\psi_0 - \phi_0)/\ell_0 \sin \phi_0$.  While, in general, the equilibrium
state of a curved raft will not have $\phi_{i\rho}=\phi_0$, $\psi_{ij\rho\sigma}=\psi_0$ and $r_{i\rho}=\ell_0$ everywhere, we may nevertheless use this relationship to 
define a proxy for 
the average curvature of the raft.  Recognizing that in numerical studies
stretching effects are small, we ignore variations in $r_{i\rho}$ and define
$
C =\sin (\langle \psi \rangle - \langle  \phi \rangle)/\ell_0 \sin(\langle  \phi \rangle)$,
where $\langle \cdot\rangle$ is an average over cells and vertices.
The colony is in the \textit{flag-in} (\textit{flag-out}) state 
when $C >0$ ($C<0$).

The simplest model of raft dynamics localizes the viscous
drag to the individual cell and vertex positions
${\bf r}_\gamma$ according
to a gradient flow $\zeta\partial_t{\bf r}_\gamma=-\partial E/\partial \mathbf{r}_\gamma$ driven by the force derived from \eqref{eq:mech_e}.
We solve this dynamics numerically with forward integration. 
Since the $\hat{\mathbf{n}}_i$ are constrained to have unit length, they are 
normalized after each step in the direction of the negative gradient, making the 
dynamical algorithm follow a projected gradient descent \cite{eicke1992iteration}.
Via a rescaling of time
we may set one of the elastic constants to unity (say, $k_\phi$) and need only
consider the ratios $K_\psi=k_\psi/k_\phi$ and $K_\ell=k_\ell/k_\phi$.

Interconversion between the \textit{flag-in} and \textit{flag-out}
states is shown in Figs. \ref{fig4}(a-d) following 
an abrupt change in the preferred angle pair $(\phi_0,\psi_0)$
that crossing the line of equality 
$\psi_0=\phi_0$ that divides the states (Fig. \ref{fig4}(e)),
as during raction/relaxation of the F-actin ring in response to a stimulus. 
The intermediate shapes exhibit a ring of inflection points similar to those
seen in experiments on {\it C. flexa} and also in the inversion the algae {\it Pleodorina} \cite{hohn2016distinct} and larger species \cite{viamontes1977cell,PLOS}.
Tracking the energy as each of the two equilibria 
is achieved, the picture that emerges in Fig. \ref{fig4}(f) is evolution 
on a double-well potential energy landscape as a biasing field is switched in sign.

We have shown that simple models can explain 
the swimming, feeding, and inversion of the recently discovered 
multicellular choanoflagellate {\it C. flexa} \cite{brunet2019light}.  
These results suggest further exploration on  
a possible continuum description of the
sheets, fluid-structure interactions during locomotion, 
dynamics of photokinesis, and developmental processes of 
these remarkable organisms.

\begin{acknowledgments}
We gratefully acknowledge Gabriela Canales and Tanner Fadero for high speed imaging assistance and 
Kyriacos Leptos for comments and suggestions.
This work was supported in part by 
a Research Fellowship from Peterhouse, Cambridge (LF),
a Churchill Scholarship (AK), 
JSPS Kakenhi (TI), The John Templeton Foundation and 
Wellcome Trust Investigator
Grant 207510/Z/17/Z (REG). For the purpose of open access, the authors have applied a CC BY public copyright license to any Author Accepted Manuscript version arising from this submission.

\end{acknowledgments}

\vfil
\eject

\begin{widetext}

\section{Supplemental Material}
This file contains additional experimental results 
on flagellar dynamics and geometry of 
{\it C. flexa}.

\setcounter{equation}{0}
\setcounter{figure}{0}
\setcounter{table}{0}
\setcounter{page}{1}
\makeatletter
\renewcommand{\theequation}{S\arabic{equation}}
\renewcommand{\thefigure}{S\arabic{figure}}
\renewcommand{\bibnumfmt}[1]{[S#1]}
\renewcommand{\citenumfont}[1]{S#1}

\bigskip

\centerline{\bf I. VIDEO IMAGING AND ANALYSIS}

\medskip

\centerline{\bf A. Supplementary Videos}

\smallskip

{\it C. flexa} sheets were imaged in FluoroDishes (World Precision Instruments FD35-100) by differential interference contrast (DIC) microscopy using a $40\times$ (water immersion, C-Apochromat, 1.1 NA) Zeiss objective mounted on a Zeiss Observer Z.1 with a pco.dimax cs1 camera. 

Flagellar characteristics reported in Table \ref{table1} were obtained as follows. Beat frequencies 
were determined by averaging over five cycles for each of twenty randomly selected cells. All other measurements are averages over ten randomly selected cells. The comparatively large wavelength for {\it flag-out} sheets may be due in part to the fact that the flagellar waveform in that state 
is not sinusoidal, and its wavelength is thus less well defined than in the {\it flag-in} state.

\begin{figure}[h]
 \includegraphics[width=0.50\linewidth]{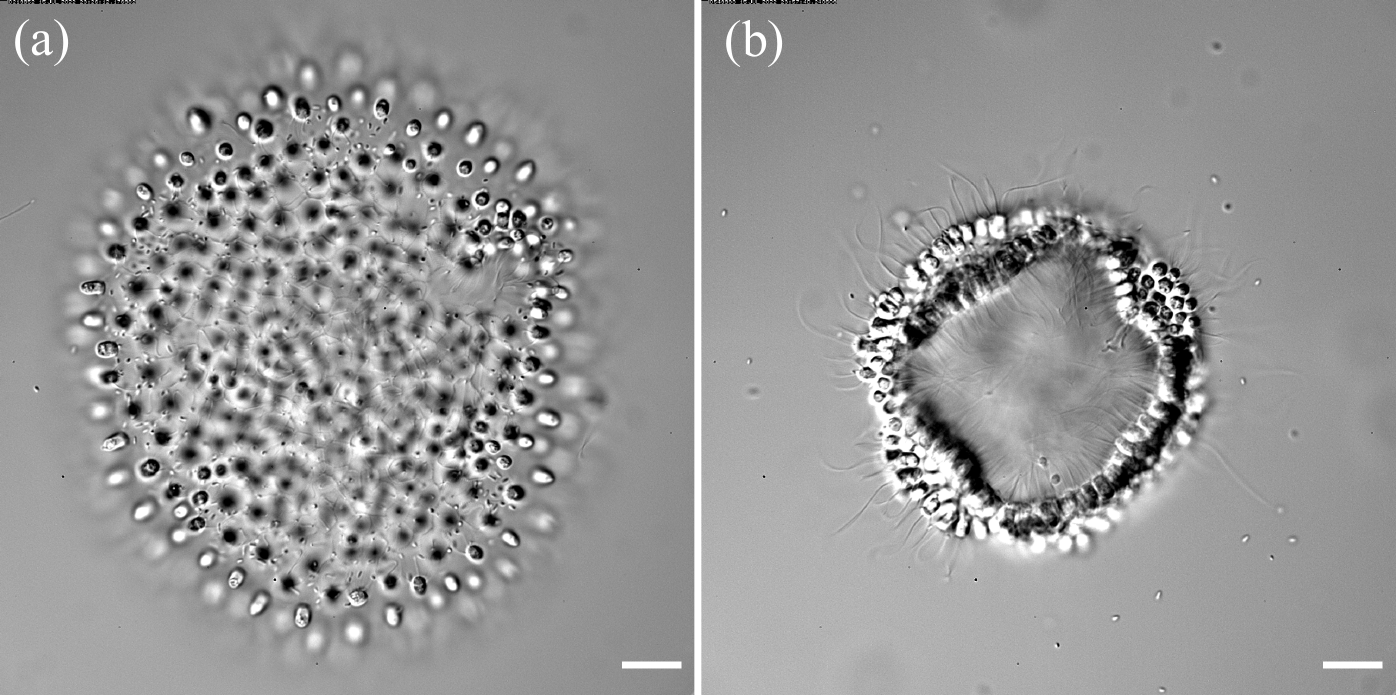}
 \caption{Flagellar dynamics in \textit{C. flexa} colonies. (a) Snapshot from Video 1, a high speed recording of
 a {\it flag-in} sheet used to determine flagellar characteristics reported in Table \ref{table1}.  Movie
 is set to play at $17\times$ slower than real time. (b) As in (a), but for the {\it flag-out} state in Video 2. 
 Scale bars are $10\,\mu$m. }
 \label{figS1}
\end{figure}

 \begin{table}[h]
 \begin{ruledtabular}
 \begin{tabular}
 {|c|c|c|c|c|}
  conformation & {beat frequency $f$} & length $2L$ &  amplitude $d$ & wavelength $\lambda$\\
  \hline
{\it flag-in} & {$45\pm 4$ Hz} &  $26 \pm 5$ $\mu$m & $2.4\pm 0.6$ $\mu$m & $9\pm 3$ $\mu$m\\
{\it flag-out} & {$43\pm 8$ Hz} &  $23\pm 4$ $\mu$m & $2.2\pm 0.7$ $\mu$m & $15 \pm 3$ $\mu$m\\
 \end{tabular}
 \end{ruledtabular}
 \caption{Measurements of flagellar characteristics for the 
 {\it flag-in} and {\it flag-out} sheets in Videos S1 and S2.  Uncertainties 
 reported are standard deviations.}
 \label{table1}
 \end{table}

\medskip

\centerline{\bf B. Estimating the propulsive force from the flagella}
\smallskip

In the fluid mechanics model of the \textit{C. Flexa} raft, we approximate the flagella beating as an effective propulsive point force $\mathbf{F}$ acting in the direction $\hat{\mathbf{n}}$. The magnitude of this force $F$ can be approximated using the resistive force theory \cite{Laugabook1} as
\begin{equation}
    F = 2L\left(\zeta_\perp-\zeta_\parallel\right) (1-\beta)  f \lambda
\end{equation}
where $2L$ is the flagella length, $f$ the beat frequency and $\lambda$ the projected wavelength in the direction of the traveling sinusoidal wave (i.e. $\hat{\mathbf{n}}$), the values of which are listed in Table \ref{table1}. Meanwhile, 
\begin{equation}
    \zeta_\perp = \frac{4\pi\mu}{\ln(2L/r)} \quad \mbox{and} \quad  \zeta_\parallel = \frac{2\pi\mu}{\ln(2L/r)}
\end{equation}
are the transverse and longitudinal drag coefficients of a cylindrical filament of radius $r$, approximated using the resistive force theory, and $\beta$ is a coefficient that depends on the flagella waveform. Although the flagella waveform is not necessarily sinusoidal, in the absence of better measurements, the value of $\beta$ is approximated, assuming the flagella takes a sinusoidal waveform $f(x)$ with wavelength $\lambda$ and amplitude $d$ (Table \ref{table1}), as
\begin{equation}
    \beta = \frac{\int_0^\lambda \left( 1+f'(x)\right)^{-1/2} dx}{\int_0^\lambda \left( 1+f'(x)\right)^{1/2} dx}, 
\end{equation}
which can be found by numerically.
In the limiting of $2 \pi d/\lambda \ll 1$, $\beta \approx 2\pi^2 (d/\lambda)^2$.

\bigskip

\centerline{\bf II. CONFOCAL IMAGING}

\medskip

Sheets in Figs. \ref{figS2} and \ref{figS3} were fixed and stained with FM1-43FX or with Alexa 488-phalloidin as in \cite{brunet2019light1}. Sheets were imaged on Zeiss LSM 880 with AiryScan using a 63x, 1.4 NA C Apo oil immersion objective (Zeiss). Z-projections were generated with Fiji \cite{schindelin}. Packing fraction was estimated by projecting cell bodies located within the same plane in a locally flat portion of the sheet and by manually outlining the border of the colonies (red dotted line in Fig. \ref{figS2}).

\begin{figure}[h]
    \includegraphics[width=0.9\linewidth]{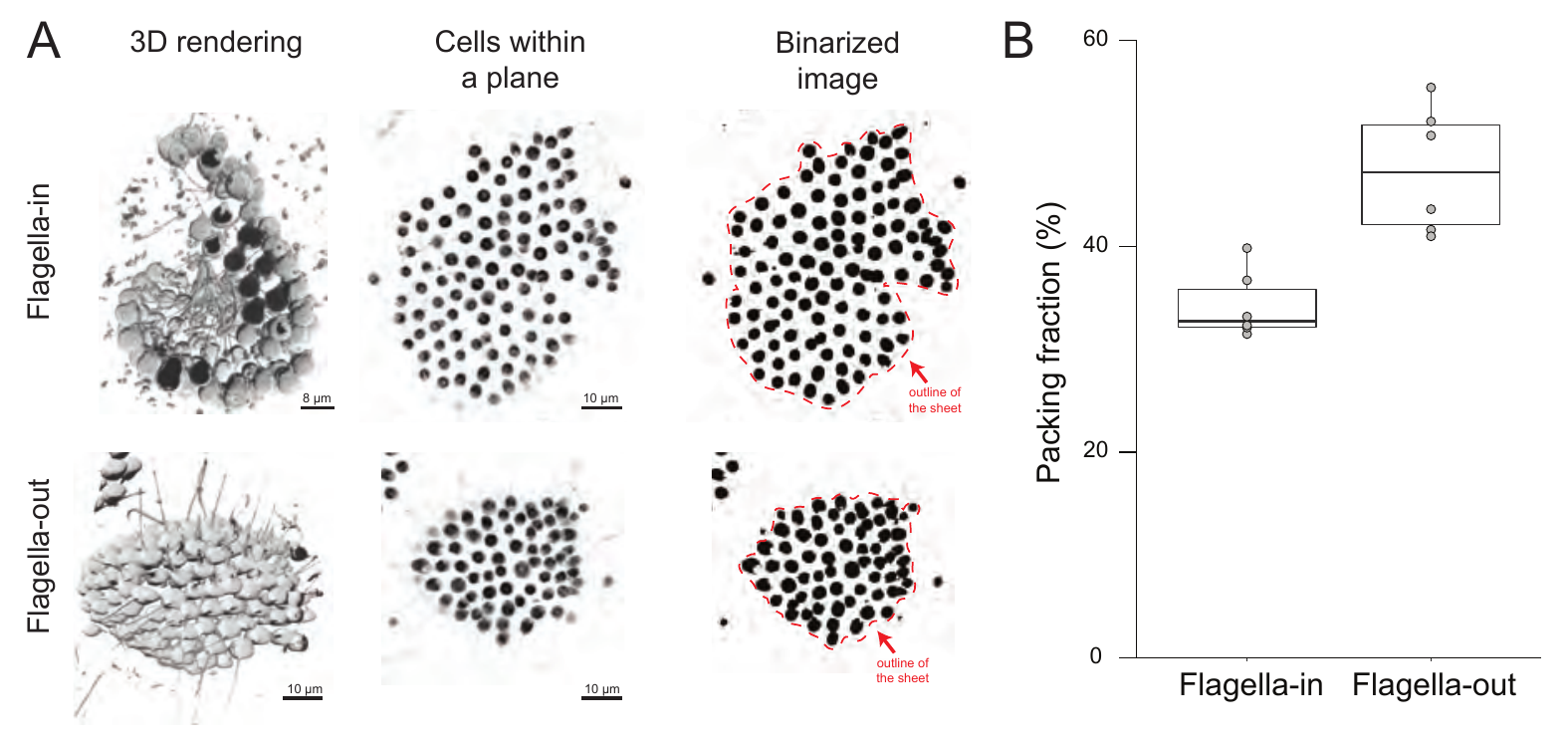}
    \caption{Packing fraction in flagella-in and flagella-out {\it C. flexa} colonies. (a) 3D stacks (left column) of sheets stained with the fluorescent membrane marker FM 1-43 FX were imaged by confocal microscopy to determine sheet morphology. Packing fraction was computed by doing a Z-projection of a locally flat portions of individual sheets (middle column) and generating a binarized image in which the area occupied by individual cells appears black (right column). Packing fraction is the ratio of the area occupied by cells to the total colony area within that plane (area within the red dotted line). (b) Boxplot depicting packing fraction values for 6 sheets with flagella out and 6 sheets with flagella-in. p=0.2\% by the Mann-Whitney test.}
    \label{figS2}
\end{figure}

 \noindent Packing fraction was then computed as the ratio of the area occupied by cells (black area in binarized image in Fig. \ref{figS2}) to the total area occupied by the colony (area within the red dotted line in Fig. \ref{figS2}). Polygonal collar borders in Fig. \ref{figS3} were manually outlined and colored with Adobe Illustrator 27.3.1 (2023). Hexagonal and pentagonal outlines were counted in 9 colonies and counts are reported in Table \ref{table2}.

\begin{figure}[t]
    \includegraphics[width=0.9\linewidth]{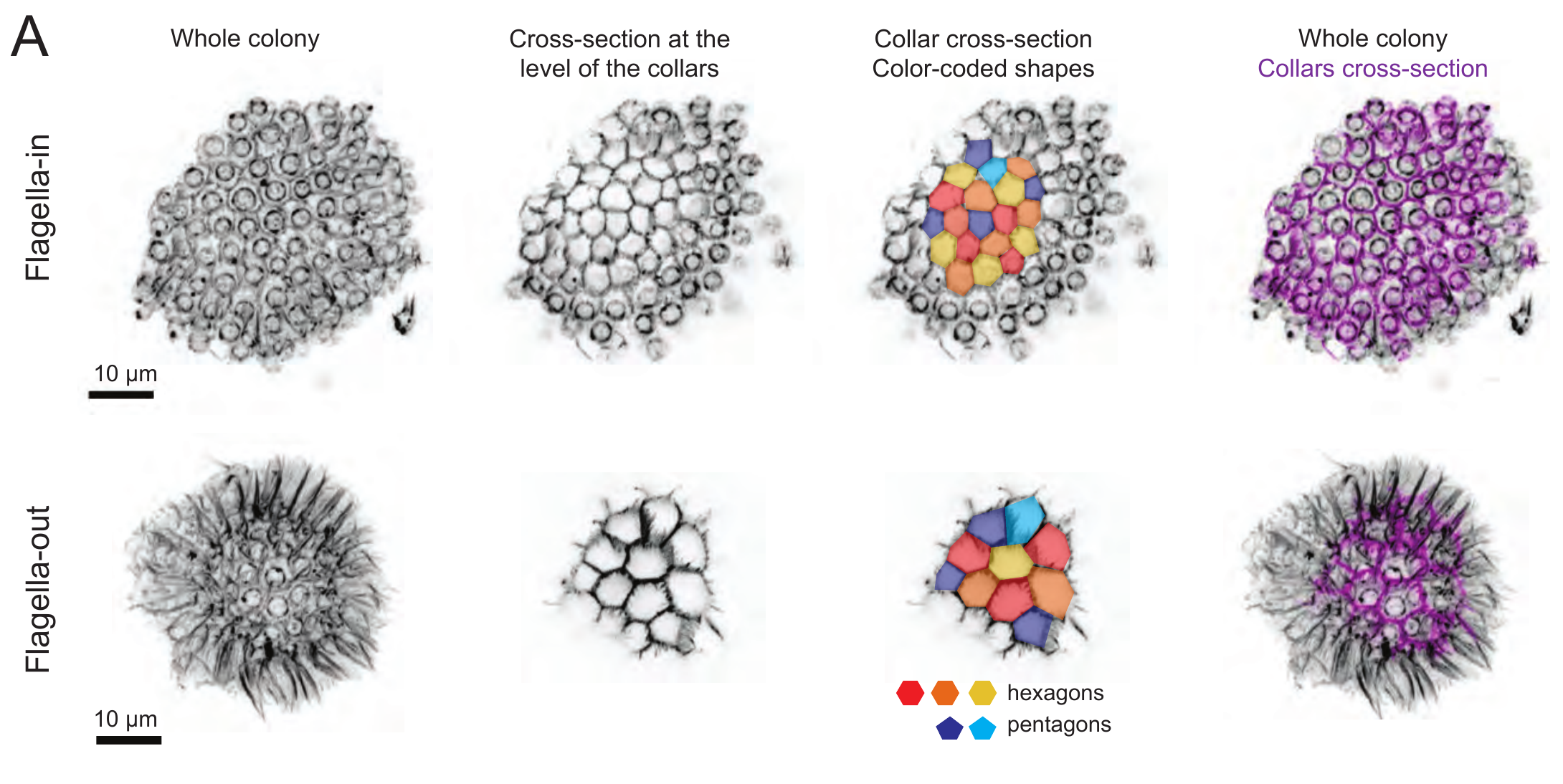}
    \caption{Hexagonal and polygonal collar outlines within flagella-in and flagella-out {\it C. flexa} colonies. F-actin within colonies was stained with fluorescent phalloidin, which outlines the microvillous collars linking cells, as well as the actin cytoskeleton within the cell body. Left column: whole-colony Z-projections. Middle left: Z-projections of a few planes intersecting many collars, showing the polygonal outlines of collar contacts. Middle right: color-coded polygons showing a majority of hexagons and a minority of pentagons. Right: whole-colony Z-projection with the plane containing most collars outlined in purple.}
    \label{figS3}
\end{figure}

\begin{table}[h]
 \caption{Numbers of hexagons and pentagons in representative sections of $9$ colonies of {\it C. flexa}.}
 \begin{ruledtabular}
 \begin{tabular}
  {|c|c|c|c|c|c|c|c|c|c|c|}
  colony \# & $1$ & $2$ & $3$ & $4$ & $5$ & $6$ & $7$ & $8$ & $9$ & total\\
  \hline
hexagons  & 4 & 4 & 2 & 8 & 8 & 15 & 6 & 6 & 8 & 61\\
pentagons & 4 & 0 & 1 & 2 & 2 & 7 & 2 & 3 & 2 & 23\\
 \end{tabular}
 \end{ruledtabular}
 \label{table2}
 \end{table}

 \vfil
 \eject

\end{widetext}

\end{document}